\documentstyle[aps,amssymb,12pt]{revtex}

\begin{document}
\title{Generalized Slow Roll Conditions and the Possibility of Intermediate Scale
Inflation in Scalar-Tensor Theory}
\author{J.R. Morris\thanks{%
E-mail: jmorris@iun.edu}}
\address{{\it Physics Dept., Indiana University Northwest,}\\
{\it 3400 Broadway, Gary, Indiana 46408}\\
\smallskip\ }
\date{}
\maketitle

\begin{abstract}
Generalized slow roll conditions and parameters are obtained for a general
form of scalar-tensor theory (with no external sources), having arbitrary
functions describing a nonminimal gravitational coupling $F(\phi )$, a
Kahler-like kinetic function $k(\phi )$, and a scalar potential $V(\phi )$.
These results are then used to analyze a simple toy model example of chaotic
inflation with a single scalar field $\phi $ and a standard Higgs potential
and a simple gravitational coupling function. In this type of model
inflation can occur with inflaton field values at an {\it intermediate}
scale of roughly $10^{11}$ GeV when the particle physics symmetry breaking
scale is approximately $1$ TeV, provided that the theory is realized within
the Jordan frame. If the theory is realized in the Einstein frame, however,
the intermediate scale inflation does not occur.

\smallskip\ 


\bigskip\ 

\newpage\ 
\end{abstract}

\section{Introduction}

Our experience with gravitation leads us to believe that Einstein gravity
provides a good description, at least at low energies. However, we can not
dismiss the possibility that some generalization, such as scalar-tensor (ST)
theory, with an appropriate low energy Einstein limit, provides a more
accurate description, especially at high energies where the deviation from
Einstein gravity can be significant. ST theories include the special cases
of Brans-Dicke theory\cite{BD} and dilaton gravity, and are physically
motivated by ideas such as Mach's principle\cite{BD}, string theory\cite
{lest},\cite{Pol} and other higher dimensional theories. When investigating
the phenomenon of inflation, it is therefore natural to study ST
modifications to Einstein gravity and their implications\cite{ST},\cite
{string}.

Here, attention is focused upon a fairly general form of ST theory
(equivalent to hyperextended scalar-tensor gravity\cite{TV}) where a single
scalar field $\phi $ couples nonminimally to gravity through the function $%
F(\phi )$ and whose dynamics is partially governed by a scalar potential $%
V(\phi )$. A kinetic function $k(\phi )$, which resembles a Kahler metric in
supersymmetric theories, allows for a noncanonical kinetic term and permits
the ST theory to be written in a form resembling generalized Brans-Dicke
gravity or dilaton gravity, for instance. The Einstein limit is realized
when $F=1$. The usual assumptions and approximations made for slow roll
inflation in the Einstein theory\cite{L} can be simply extended to
accommodate the functions $F$, $k$, and $V$ in the ST theory with the
imposition of a very simple and mild requirement that each function (or its
inverse) have a sufficiently rapid convergence of its Taylor series power
expansion so that it can be well approximated with a finite number $N$ of
terms with $N\ll H/(\dot{\phi}/\phi )$. With these conditions, the usual
slow roll approximations can be implemented to obtain the slow roll
equations of motion (EOM) for the field $\phi $ in a flat Robertson-Walker
spacetime. The slow roll parameters can be defined for the ST theory in
terms of the functions $F$, $k$, and $V$. (However, an ``effective
potential'' $U(\phi )=V/F^2$ arises in the ST theory and appears in the EOM,
and the parameters can be written in a more economical form by using the
function $U$.) For a canonical kinetic term ($k=1$) in the Einstein limit ($%
F=1$), the slow roll parameters and EOM reduce to the usual ones.

Applying the generalized results to slow roll chaotic inflation, we explore
the possibility that ST modifications can allow chaotic inflation\cite{Lin}
to occur at {\it intermediate} energy scales, characterized by inflaton
field values $\phi \ll M_P$, that are far below the Planck scale and
therefore radically different from those normally assumed. This feature also
appears in other models with minimally coupled scalar fields that have
recently been suggested, such as ``supernatural inflation''\cite{RSG} or
``assisted inflation''\cite{lms} in higher dimensional theories\cite{KO}.
However, these inflationary models involve more than one scalar field, as
Randall, Soljacic, and Guth\cite{RSG} have argued must be the case. But,
within the context of ST theory, the nonminimal coupling of a single
inflaton to gravity can have an important effect, allowing inflation to
proceed at an intermediate energy scale $E\ll M_P$ for a class of coupling
functions $F(\phi )$. This result stems from an interesting possibility of a
relative increase in the strength of gravity at an intermediate scale in the
ST theory, with the nonminimal coupling $F(\phi )$ giving rise to an
effective gravitational coupling $\bar{\kappa}^2(\phi )=\kappa ^2/F(\phi )$
that is an {\it increasing} function of the inflaton field $\phi $. This
effect can allow enough inflation to occur for a relatively small range of
values of $\phi $ at a relatively low energy scale.

Specifically, the possibility of an intermediate scale inflation is studied
within the context of a toy model with a standard type of quartic Higgs
potential $V(\phi )$ and a simple nonminimal scalar coupling function $%
F(\phi )$ that is a {\it decreasing} function of $|\phi |$. The Higgs
potential locates the vacuum state at $|\phi |=\eta $, where $\eta $ is
assumed to be small in comparison to the Planck mass $M_P$. The function $F$
becomes unity in the vacuum, i.e. $F(\eta )=1$, so that the nonminimal
coupling approaches minimal coupling in the vacuum. From the slow roll
conditions, along with the requirement that there be enough $e$-folds of
inflation, the final and initial values of $\phi $ during inflation can be
determined. Taking $\eta \approx 1$ TeV, as an example, we find inflation
occurring for $\phi \sim 10^{11}$ GeV, i.e., at an intermediate scale. The
possibility of intermediate scale inflation exists with the proviso that the
scalar-tensor theory is physically realized within the {\it Jordan}
conformal frame, where there is an explicit nonminimal coupling of the
scalar field. Of course, this explicit nonminimal coupling can be removed by
performing a conformal transformation to the {\it Einstein} frame. However,
only one conformal frame can be the physical frame, and we find that
intermediate scale inflation does not occur if the Einstein frame is
physically realized.

The action for the scalar-tensor theory is presented and the field equations
and cosmological EOM are obtained in section II. The usual slow roll
conditions and approximations are generalized in section III and applied to
obtain the slow roll EOM and slow roll parameters. A toy model of
intermediate scale chaotic inflation is presented and analyzed in section
IV. In section V we mention the debate as to whether it should be the Jordan
frame or the Einstein frame that should be regarded as being the physical
frame, and show that the intermediate scale inflation of the toy model does
not occur in the Einstein frame. A short summary and some concluding remarks
are offered in section V.

\section{Scalar-Tensor Theory}

We consider scalar-tensor (ST) theory formulated in the Jordan conformal
frame, where there is an explicit nonminimal coupling, described by the
function $F(\phi )$, between the gravitational field and the scalar field $%
\phi $. The action can be written in the form\footnote{%
We use a metric with signature $(+,-,-,-)$ and a Ricci tensor $R_{\mu \nu
}=\partial _\nu \Gamma _{\mu \lambda }^\lambda -\partial _\lambda \Gamma
_{\mu \nu }^\lambda -\Gamma _{\mu \nu }^\rho \Gamma _{\rho \sigma }^\sigma
+\Gamma _{\mu \sigma }^\rho \Gamma _{\nu \rho }^\sigma $.} 
\begin{equation}
\begin{array}{ll}
S & =%
\displaystyle \int 
d^4x\sqrt{-g}\left\{ F(\phi )%
{\displaystyle {R \over 2\kappa ^2}}
+\frac 12k(\phi )(\partial \phi )^2-V(\phi )\right\}
\end{array}
,  \label{e1}
\end{equation}

\noindent where $\kappa ^2=8\pi G$, and the nonminimal coupling function $F$%
, the kinetic function $k$, and the scalar potential $V$ are arbitrary
functions of $\phi $ only. No metric dependent external source terms have
been included in the action, but a dilaton coupled cosmological constant can
be accommodated by the function $V(\phi )$. For a canonical kinetic term $%
k(\phi )=1$, whereas for pure Brans-Dicke theory, we could write $F(\phi )=%
{\displaystyle {\kappa ^2 \over 8\pi }}
\phi $, $k(\phi )=\omega /(8\pi \phi )$, and $V(\phi )=0$. With different
parametrizations of the functions $F$, $k$, and $V$, we can write the action
(\ref{e1}) in the following forms:

$\bullet $ {\it Generalized Brans-Dicke theory: }Taking $F(\phi )=%
{\displaystyle {\kappa ^2 \over 8\pi }}
\phi $ and $k(\phi )=\omega (\phi )/(8\pi \phi )$, the action for a
generalized Brans-Dicke theory can be written in the form 
\begin{equation}
S_{BD}=\frac 1{16\pi }\int d^4x\sqrt{-g}\left\{ \phi R+\frac{\omega (\phi )}%
\phi (\partial \phi )^2-16\pi V(\phi )\right\} .  \label{e2}
\end{equation}

$\bullet $ {\it Dilaton gravity}: For dilaton gravity we can write $F(\phi
)=e^{-\phi }$, $k(\phi )=-%
{\displaystyle {F(\phi ) \over \kappa ^2}}
=-%
{\displaystyle {e^{-\phi } \over \kappa ^2}}
$, and $V(\phi )=%
{\displaystyle {e^{-\phi } \over 2\kappa ^2}}
W(\phi ) $. In this case the action (\ref{e1}) can be alternatively written
in the form 
\begin{equation}
S_{DG}=\frac 1{2\kappa ^2}\int d^4x\sqrt{-g}e^{-\phi }\left\{ R-(\partial
\phi )^2-W(\phi )\right\} .  \label{e3}
\end{equation}

$\bullet $ {\it Hyperextended scalar-tensor gravity: }The case of
hyperextended scalar-tensor gravity (HSTG) studied by Torres and Vucetich%
\cite{TV}, and Torres\cite{T}, can be described with $F(\phi )=%
{\displaystyle {\kappa ^2 \over 8\pi G(\phi )}}
$, $k(\phi )=%
{\displaystyle {\omega (\phi ) \over 8\pi \phi }}
$, where $G(\phi )$ is an arbitrary function of $\phi $, with the action
assuming the form 
\begin{equation}
S_{HSTG}=\frac 1{16\pi }\int d^4x\sqrt{-g}\left\{ \frac R{G(\phi )}+\frac{%
\omega (\phi )}\phi (\partial \phi )^2-16\pi V(\phi )\right\} .  \label{e4}
\end{equation}

\noindent The action in (\ref{e1}) is equivalent to the action in (\ref{e4}%
), where there are three free functions.

The variation of the action in (\ref{e1}) gives the field equations

\begin{equation}
\begin{array}{ll}
G_{\mu \nu }= & -%
{\displaystyle {F_{,\phi } \over F}}
\nabla _\mu \partial _\nu \phi -%
{\displaystyle {1 \over F}}
\left[ F_{,\phi \phi }+\kappa ^2k\right] \partial _\mu \phi \partial _\nu
\phi \\ 
& +%
{\displaystyle {g_{\mu \nu } \over F}}
\left\{ \left[ F_{,\phi \phi }+%
{\displaystyle {\kappa ^2k \over 2}}
\right] (\partial \phi )^2+F_{,\phi }\Box \phi -\kappa ^2V\right\} ,
\end{array}
\label{e5}
\end{equation}

\begin{equation}
\kappa ^2k\Box \phi +\frac 12\kappa ^2k_{,\phi }(\partial \phi )^2+\kappa
^2V_{,\phi }-\frac 12F_{,\phi }R=0.  \label{e6}
\end{equation}

\noindent where $G_{\mu \nu }=R_{\mu \nu }-\frac 12g_{\mu \nu }R$ is the
Einstein tensor. Taking the trace of (\ref{e5}) to obtain an expression for
the Ricci scalar $R$ and inserting into (\ref{e6}) gives 
\begin{equation}
\left[ \frac{3(F_{,\phi })^2}{2F}+\kappa ^2k\right] \Box \phi +\left\{ \frac{%
F_{,\phi }}{2F}\left[ 3F_{,\phi \phi }+\kappa ^2k\right] +\frac 12\kappa
^2k_{,\phi }\right\} (\partial \phi )^2+\kappa ^2F^2U_{,\phi }=0,  \label{e7}
\end{equation}

\noindent where $F_{,\phi }=\partial F/\partial \phi $, $F_{,\phi \phi
}=\partial ^2F/\partial \phi ^2$, etc. and 
\begin{equation}
U=\frac V{F^2}  \label{e8}
\end{equation}

\noindent is an effective potential induced by the scalar curvature $R$. We
shall take the field equations to be given by (\ref{e5}) and (\ref{e7}).

To obtain the cosmological equations of motion (EOM), we adopt the metric
for a flat Robertson-Walker spacetime, 
\begin{equation}
ds^2=dt^2-a^2(t)d\vec{x}\cdot d\vec{x}.  \label{e9}
\end{equation}

\noindent Eqs. (\ref{e5}), (\ref{e7}), and (\ref{e9}) then give the EOM 
\begin{equation}
H^2=\frac{\kappa ^2}{3F}\left[ \frac 12k\dot{\phi}^2+V\right] -\frac{%
F_{,\phi }}FH\dot{\phi},  \label{e10}
\end{equation}
\begin{equation}
\begin{array}{l}
\left[ 
{\displaystyle {3(F_{,\phi })^2 \over 2F}}
+\kappa ^2k\right] (\ddot{\phi}+3H\dot{\phi})+\left\{ 
{\displaystyle {F_{,\phi } \over 2F}}
\left[ 3F_{,\phi \phi }+\kappa ^2k\right] +\frac 12\kappa ^2k_{,\phi
}\right\} \dot{\phi}^2 \\ 
+\kappa ^2F^2U_{,\phi }=0,
\end{array}
\label{e11}
\end{equation}

\noindent where $H=\dot{a}/a$ is the Hubble parameter. Notice that there is
an effective gravitational coupling $\bar{\kappa}$ defined by $\bar{\kappa}%
^2=\kappa ^2/F$.

\section{Generalized Slow Roll Conditions and Parameters}

In most inflationary models, inflation takes place under the slow roll
conditions\cite{L} that (i) the inflaton field $\phi $ evolves slowly in
comparison to the expansion rate of the universe, and (ii) the kinetic
energy density of the inflaton is negligible in comparison to the potential
energy density, i.e., the inflation is driven by the potential. These
conditions, which can be stated more quantitatively as 
\begin{equation}
|\ddot{\phi}|\ll H|\dot{\phi}|\ll H^2|\phi |,  \label{e20}
\end{equation}
\begin{equation}
\frac 12|k(\phi )|\,\dot{\phi}^2\ll |V(\phi )|,  \label{e21}
\end{equation}

\noindent will be assumed to hold during an inflationary epoch. However, as
pointed out by Torres\cite{T}, the condition given by (\ref{e20}) can be
generalized to read $|\ddot{f}|\ll H|\dot{f}|\ll H^2|f|$, for an arbitrary
function $f(\phi )$, provided that the function $f$ has a power series
expansion that converges sufficiently fast. (Actually, this condition is 
{\it sufficient}, but not {\it necessary}, as we will see shortly.) To be
more specific, let us assume that $f(\phi )$ has a power series
representation of the form $f=\sum\limits_{n=0}^\infty f_n$, with $%
f_n=c_n\phi ^n$ and a nonzero radius of convergence. Let us also assume that
the series converges fast enough that $f$ can be well approximated by $%
f\approx \sum\limits_{n=0}^Nf_n$, with $N\ll H/(\dot{\phi}/\phi )$; i.e.,
the remainder is small in comparison to $f$, and the integer $N$ is not too
terribly large. We then have that $|\dot{f}_n/f_n|=n|\dot{\phi}/\phi |\ll H$%
, so that each term in the expansion of $\dot{f}$ is small compared to the
corresponding term in the expansion for $f$, from which we conclude that $|%
\dot{f}|\ll H|f|$. Furthermore, for the acceleration of the function $f(\phi
)$, we have $|\ddot{f}_n/\dot{f}_n|=|(n-1)\dot{\phi}/\phi +\ddot{\phi}/\dot{%
\phi}|$, and by (\ref{e20}) each of these terms on the right hand side is
small compared to $H$, allowing us to conclude that $|\ddot{f}|\ll H|\dot{f}%
| $. Therefore, (\ref{e20}) can be generalized to read 
\begin{equation}
|\ddot{f}|\ll H|\dot{f}|\ll H^2|f|  \label{e22}
\end{equation}

\noindent whenever the function $f(\phi )$ has a sufficiently rapidly
convergent power series Taylor expansion. Actually, this convergence
assumption can be relaxed somewhat. For instance, even if a power series
expansion for $f$ does not converge within some domain, but $g=1/f$ does
converge with sufficient rapidity, then $|\ddot{g}|\ll H|\dot{g}|\ll H^2|g|$
implies that (\ref{e22}) is satisfied for the function $f$. (As an example, $%
f=(1+c\phi ^2)^{-1}$, where $c$ is a constant, does not have a convergent
Maclaurin series expansion for $|c|\phi ^2>1$, but $g=f^{-1}$does, and
therefore both $g$ and $f$ satisfy the condition (\ref{e22}) whenever the
condition (\ref{e20}) is satisfied.) To obtain our generalized slow roll
EOM, it will be assumed that we are dealing with functions $F$, $F_{,\phi }$%
, $k$, and $U$ (or $V$) that satisfy (\ref{e22}).

Applying the above slow roll conditions to (\ref{e10}) and (\ref{e11}) (see
the appendix for details) gives the slow roll EOM 
\begin{equation}
H^2=\frac{\kappa ^2}{3F}V,  \label{e22a}
\end{equation}
\begin{equation}
3Hk\dot{\phi}+F^2U_{,\phi }=0.  \label{e22b}
\end{equation}

We now define the slow roll parameters 
\begin{equation}
\varepsilon _{SR}=\frac 1{2\kappa ^2}\left[ \frac Fk\left( \frac{U_{,\phi }}U%
\right) ^2\right] ,  \label{p1}
\end{equation}
\begin{equation}
\eta _{SR}=\frac 1{\kappa ^2}\left[ \frac 1{FU}\frac \partial {\partial \phi 
}\left( \frac{F^2U_{,\phi }}k\right) \right] .  \label{p2}
\end{equation}

\noindent  For the case of minimal coupling ($F=1$) and a canonical kinetic
term ($k=1$) these parameters collapse to the usual expressions. The
condition given by (\ref{e21}), when used in conjunction with (\ref{e22a})
and (\ref{e22b}), then translates into the condition $\left| 
{\displaystyle {\frac 12k\dot{\phi}^2 \over V}}
\right| =%
{\displaystyle {|\varepsilon _{SR}| \over 3}}
\ll 1$. (The condition $|\varepsilon _{SR}|\ll 1$ also follows from the
condition $|\dot{U}/U|\ll H$.) Next, we use the fact that an application of (%
\ref{e22}) to the functions $F(\phi )$ and $U(\phi )$ implies that $|\dot{H}%
/H|\ll 1$. Then taking the time derivative of $\dot{\phi}$, using (\ref{e22b}%
), and applying (\ref{e22a}), the condition that $|\ddot{\phi}/(H\dot{\phi}%
)|\ll 1$ translates into the condition $|\eta _{SR}|\ll 1$. The slow roll
conditions can then be stated in terms of the parameters $\varepsilon _{SR}$
and $\eta _{SR}$: 
\begin{equation}
|\varepsilon _{SR}|\ll 1,\,\,\,\,\,\,\,\,\,\,|\eta _{SR}|\ll 1.  \label{p3}
\end{equation}

\noindent The violation of these conditions signals the end of the
inflationary period.

The onset of inflation occurs at a time $t_i$ when the inflaton has a value $%
\phi =\phi _i$ and inflation ends at a time $t_f$ when $\phi =\phi _f$. The
amount of inflation is given by the number $N\approx \int_{t_i}^{t_f}Hdt$ of 
$e$-folds of the scale factor, which from the slow roll EOM is 
\begin{equation}
N\approx \int_{t_i}^{t_f}\frac{H^2\dot{\phi}}{H\dot{\phi}}dt=%
\displaystyle \int 
_{\phi _i}^{\phi _f}\frac{H^2}{H\dot{\phi}}d\phi \approx -\kappa ^2%
\displaystyle \int 
_{\phi _i}^{\phi _f}\left[ \frac kF\left( \frac U{U_{,\phi }}\right) d\phi
\right] .  \label{p4}
\end{equation}

\section{Intermediate Scale Chaotic Inflation}

We have seen that the slow roll inflationary conditions and parameters for a
scalar-tensor theory, where the inflaton is nonminimally coupled to gravity
through the function $F(\phi )$, are modified from the usual conditions for
a minimally coupled inflaton field. In other words, the conditions under
which inflation begins and ends are controlled not just by the potential $V$%
, but also by the coupling function $F$. (Here, we will consider the case of
a canonical kinetic term, $k=1$.) It therefore seems plausible that the ST
theory can accommodate a class of coupling functions $F(\phi )$ that would
allow the onset of chaotic inflation to appear when the inflaton acquires a
value $\phi \ll M_P=G^{-1/2}$, well below the Planck scale.

In the usual case of a minimally coupled inflaton (and no other scalar
fields), we typically find that inflation begins and ends for $\phi \gtrsim
M_P$, as we can easily see from the following example. Consider a minimally
coupled theory with $F=1$, $k=1$, and a quartic Higgs potential $V=\lambda
(\phi ^2-\eta ^2)^2=\lambda \eta ^4(\bar{\phi}^2-1)^2$, where $\bar{\phi}%
=\phi /\eta $, and the mass scale, determined by the parameter $\eta $ in
the potential, comes purely from the particle physics. At large values of $%
\phi $, i.e. $\bar{\phi}\gg 1$, we have a simple power behavior for the
potential, $V\approx \lambda \eta ^4\bar{\phi}^4$. The end of the
inflationary period is signalled by $\varepsilon _{SR}\sim 1$, which implies
that $\bar{\phi}_f\sim 4/(\kappa \eta )$, or $\phi _f\sim 4M\sim M_P$, where
we define the mass $M=\kappa ^{-1}=M_P/\sqrt{8\pi }$. The onset of
inflation, say about 70 $e$-folds earlier, occurs at a value $\phi _i$,
determined by $N\approx \kappa ^2\int_{\phi _f}^{\phi _i}\frac V{V_{,\phi }}%
d\phi $, which gives $\phi _i\sim \sqrt{8N}(\kappa \eta )^{-1/2}\approx
\,5M_P$ for $N\approx 70$. Actually, a condition on the vacuum density can
be obtained from the slow roll EOM and approximations. From the motion
equations for $H$ and $\dot{\phi}$, we have $|F^2U_{,\phi }|=|3Hk\dot{\phi}%
|\ll 3H^2|k\phi |=\kappa ^2|k\phi FU|$, which implies that 
\begin{equation}
\left| \frac{k\phi U}{FU_{,\phi }}\right| \gg M^2  \label{e26}
\end{equation}

\noindent which, for the present example, gives $\phi _i\gg 4\left( \frac M%
\eta \right) M\gg M_P$ for $\eta /M\ll 1$, implying that there are many more 
$e$-folds of inflation than the minimal number required. If $\eta \lesssim M$%
, i.e. if the symmetry breaking takes place a little below the Planck scale
(or if there is no symmetry breaking, as in $m^2\phi ^2$ or $\lambda \phi ^4$
chaotic inflation models), then we have the condition $\phi \gg M_P$, but if
the symmetry breaking takes place at a low energy scale, the condition on
the vacuum density implies that $\phi _i$ is absurdly large.

\subsection{Toy Model Example}

Here we consider a very simple toy model using the same potential $V=\lambda
\eta ^4(\bar{\phi}^2-1)^2$, $k=1$, but for the coupling function we choose
the simple function $F=%
{\displaystyle {2 \over (\bar{\phi}^2+1)}}
$, with $F\rightarrow 1$ as $\phi \rightarrow \eta $. The basic idea is to
investigate how the function $F$ can alter the inflationary conditions to
give an inflationary period at a scale $\phi \ll M$. To get plenty of $e$%
-folds over a sufficiently small range of $\phi $ values, we want $H$ to be
sufficiently large, or from (\ref{p4}), we want $|U/FU_{,\phi }|$ to be
sufficiently large. Now, if within the context of a scalar-tensor theory we
have an effective gravitational coupling $\bar{\kappa}=F^{-1/2}\kappa $
which is an {\it increasing} function of $\phi $, i.e. $F$ is a {\it %
decreasing} function of $\phi $, then for a given value of $\phi $, $H^2=%
\bar{\kappa}^2V/3>\kappa ^2V/3$, giving a {\it larger} value of $H$ than in
the minimally coupled case. This may allow inflation to proceed at a {\it %
smaller} value of $\phi $ than in the usual scenario. Naively, if in the
usual case we have inflation proceeding at $\phi \sim M$, we expect that in
the ST model we would have $\phi \sim \bar{\kappa}^{-1}=\bar{M}=F^{1/2}M$.
Therefore we expect $\phi \ll M$ if $F(\phi )\ll 1$. For the simple function 
$F=2/(\bar{\phi}^2+1)\approx 2/\bar{\phi}^2$ for $\bar{\phi}\gg 1$, the
condition $\phi \sim F^{1/2}M$ gives $\bar{\phi}\sim (\kappa \eta )^{-1/2}$,
or $\phi \sim (\eta M)^{1/2}\ll M$ for $\eta /M\ll 1$. For example, if we
take $\eta \approx 1$ TeV and $M\approx 10^{18}$ GeV, we find $\phi \sim
3\times 10^{10}$ GeV, implying inflation at an intermediate scale.

To check this qualitative argument, we can use the slow roll conditions
obtained above. Because we have simple functions, which for $\bar{\phi}\gg 1$
behave as simple powers, the estimates are easy to get. Taking the end of
inflation to coincide with $\varepsilon _{SR}\sim 1$, we find $\bar{\phi}%
_f\sim 3(\kappa \eta )^{-1/2}$, or $\phi _f\sim 3(\eta M)^{1/2}\approx
10^{11}$ GeV. We estimate $\phi _i$ from the $e$-fold formula to get $\bar{%
\phi}_i\lesssim 8(\kappa \eta )^{-1/2}$, or $\phi _i\lesssim 8(\eta
M)^{1/2}\approx 2.4\times 10^{11}$ GeV. These results agree with our
qualitative estimate. The condition on the vacuum density $|\phi
U/(FU_{,\phi })|\gg M^2$ in this case gives $\phi _i\gtrsim 3(\eta
M)^{1/2}\approx 10^{11}$ GeV. This can be compared to the condition in the
previous (minimally coupled) case where $\phi _i\gg (M/\eta )M_P\approx
10^{15}M_P$ (for $\eta \approx 1$ TeV), an enormous value! In a minimally
coupled model of inflation (without additional source terms), it appears
difficult to have inflation associated with low scale symmetry breaking.

We conclude that the scale at which inflation occurs can be substantially
modified by a nonminimal coupling of the scalar field. For a symmetry
breaking scale $\eta \approx 1$ TeV, the scale for the inflaton field $\phi $
and the {\it effective} Planck scale $\bar{M}_P=\sqrt{8\pi }\bar{M}=(\sqrt{%
8\pi /}\bar{\kappa})\sim (8\pi \eta M)^{1/2}$ are both brought down to an 
{\it intermediate} scale.\footnote{%
Note that, although inflation may occur for $\phi \sim \bar{M}=\bar{\kappa}%
^{-1}$, the {\it energy density} of the inflaton field $\sim \lambda \phi
^4\sim \lambda \bar{M}^4$ may still be small compared to $\bar{M}^4$
provided that $\lambda $ is small, so that the {\it energy} scale for
inflation can still be below the effective Planck scale.} Another way to
state this is that the {\it effective} dimensionless gravitational
``coupling constant'' $\bar{\alpha}_g=(\bar{\kappa}^2/8\pi )E^2$ approaches
unity at an energy scale of $E\sim \sqrt{8\pi }\,\bar{\kappa}^{-1}\sim (8\pi
\eta M)^{1/2}$. Therefore, if reality is described by a scalar-tensor theory
with a {\it nonminimal} coupling of the scalar field which is a {\it %
decreasing} function of $\phi $, and the scalar field is associated with a
low energy symmetry breaking, then it is possible for nontrivial
gravitational effects to begin showing up at an {\it intermediate} scale,
well below the Planck scale, which is an intriguing prospect.

Finally, it should be pointed out that the particular toy model used here as
an example, while computationally convenient, is oversimplified and not
quite realistic, in that it does not satisfy observational constraints. In
the near-vacuum sector, where $\bar{\phi}\approx 1$, we have $V(\phi
)\approx 0$ and the Lagrangian takes an approximate Brans-Dicke form 
\begin{equation}
\frac L{\sqrt{-g}}\approx \frac \Phi {16\pi }R+\frac{\kappa ^2}{16\pi }%
\left[ \frac F{(F_{,\phi })^2}\right] \frac{(\partial \Phi )^2}\Phi ,
\label{e28}
\end{equation}

\noindent where $\Phi =(8\pi /\kappa ^2)F(\phi )$ is a Brans-Dicke field,
and we can identify the Brans-Dicke parameter 
\begin{equation}
\omega \approx \kappa ^2\left| \frac F{(F_{,\phi })^2}\right| _{\phi =\eta }.
\label{e29}
\end{equation}

\noindent For our toy model with a simple power function for $F$, $\omega
\approx (\kappa \eta )^2\ll 1$, which violates the observational constraint $%
\omega \gtrsim 500$. For a realistic model we should therefore have 
\begin{equation}
\left| \frac{F_{,\phi }}{F^{1/2}}\right| _{\phi =\eta }\approx \frac \kappa {%
\omega ^{1/2}}\lesssim \frac \kappa {\sqrt{500}},  \label{e30}
\end{equation}

\noindent implying that the function $F(\phi )$ should be quite flat near $%
\phi \approx \eta $ and then decrease for $\phi \gg \eta $ where inflation
occurs.

For the case of a minimally coupled inflaton, it has been shown how the
inflaton potential can be reconstructed from slow roll parameters (see, e.g.,%
\cite{ms,sm}). This type of approach of reconstructing the form of the
theory has been extended to scalar-tensor theory by Boisseau,
Esposito-Farese, Polarski, and Starobinsky\cite{bois}, who have shown how
both of the functions $F(\phi )$ and $V(\phi )$ can be determined (with a
rescaling of $\phi $ to set $k=1$) from future observations.

\section{Jordan and Einstein Conformal Frames}

Slow roll conditions and the possibility of intermediate scale inflation in
scalar-tensor theory have been presented here in the {\it Jordan} frame,
where the scalar field $\phi $ is {\it nonminimally} coupled to the
gravitational field. However, a conformal transformation allows the model to
be recast in the {\it Einstein} frame, where the scalar field $\phi $ is 
{\it minimally} coupled to gravity. Of course, only one of these frames is
the physical frame, but there has existed quite a lot of debate and variance
in the literature as to exactly {\it which} frame should be considered the
physical frame. (For a review of this situation, see, for example, \cite{fgn}
and references therein.) The Jordan frame is taken as the physical frame in
many studies of scalar-tensor cosmology ({\it e.g.,} ref.\cite{bois}),
including studies of string cosmology ({\it e.g.}, ref.\cite{string}). The
low energy scalar-tensor field theoretic action for string theory, in the
string frame, follows from the original action for the strings, and this
frame is often considered to provide the physical interpretation, with the
strings ``seeing'' the string (Jordan) frame, not the conformally related
Einstein frame. Also, in many forms of scalar-tensor theory, the weak
equivalence principle is violated in the Einstein frame, due to the
anomalous coupling of the dilaton to matter\cite{fgn}. On the other hand,
some objections have been raised against the consideration of the Jordan
frame as the physical frame in ST models\cite{fgn}, especially those having
a negative kinetic energy term for the dilaton field at the tree level. In
addition, Torres, Schunck, and Liddle\cite{tsl} have pointed out that
whereas several mass definitions for boson stars in a scalar-tensor theory
lead to different results in the Jordan frame, they all coincide in the
Einstein frame, leading to the suspicion that the Einstein frame must be
regarded as the physical frame. (For the case of a massless scalar field,
Damour and Nordtvedt\cite{dn} have shown that general relativity acts as a
cosmological attractor of scalar-tensor theories, so that they may become
indistinguishable after early times.)

If the Jordan frame is the physical frame, i.e. the frame in which the
physical parameters of the theory coincide with those that are measured in
experiments (see, {\it e.g.}, ref.\cite{bois}), we see the possibility
arising for intermediate scale inflation for a certain class of coupling
functions $F(\phi )$ and a relatively low energy ($\eta \ll M_P$) symmetry
breaking scale from the particle physics sector. However, it should be
pointed out that our toy model considered previously does {\it not} lead to
intermediate scale inflation in the {\it Einstein} conformal frame. A
conformal transformation to the Einstein frame can be accomplished with the
rescaling $g_{\mu \nu }\rightarrow \hat{g}_{\mu \nu }$, where 
\begin{equation}
\hat{g}_{\mu \nu }=F(\phi )g_{\mu \nu }.  \label{b1}
\end{equation}

\noindent With this conformal rescaling the action (\ref{e1}) now takes the
form 
\begin{equation}
\begin{array}{ll}
\hat{S} & =%
\displaystyle \int 
d^4x\sqrt{-\hat{g}}\left\{ 
{\displaystyle {\hat{R} \over 2\kappa ^2}}
+\frac 12\hat{k}(\phi )(\hat{\partial}\phi )^2-U(\phi )\right\} ,
\end{array}
\label{b2}
\end{equation}

\noindent where $\hat{R}$ is built from the Einstein metric $\hat{g}_{\mu
\nu }$, $(\hat{\partial}\phi )^2=\hat{g}^{\mu \nu }\partial _\mu \phi
\partial _\nu \phi $, and 
\begin{equation}
\,\hat{k}=%
{\displaystyle {1 \over \kappa ^2}}
\left[ 
{\displaystyle {3(F_{,\phi })^2 \over 2F^2}}
+%
{\displaystyle {\kappa ^2k \over F}}
\right] .  \label{b3}
\end{equation}

The field equations and cosmological EOM can be simply obtained from those
presented for the Jordan frame by making the replacements $F\rightarrow \hat{%
F}=1$, $k\rightarrow \hat{k}$, $V\rightarrow U$, along with a rescaling of
the time coordinate $t\rightarrow \hat{t}$ defined by $d\hat{t}=F^{1/2}dt$
and a rescaling of the scale factor $a\rightarrow \hat{a}=F^{1/2}a$, so that
the metric takes the RW form in the Einstein frame: 
\begin{equation}
d\hat{s}^2=\hat{g}_{\mu \nu }dx^\mu dx^\nu =F[dt^2-a^2(t)d\vec{x}^2]=d\hat{t}%
^2-\hat{a}^2(\hat{t})d\vec{x}^2.  \label{b4}
\end{equation}

When the toy model of the previous section is analyzed within the Einstein
frame, intermediate scale inflation disappears, since a field redefinition $%
\phi \rightarrow \varphi $ of the scalar field is possible to obtain a
canonical kinetic term. This resets the scale of inflation $\phi _i\approx
10^{11}$GeV back to the usual Planck scale; roughly, we have $\varphi _i\sim 
\hat{k}^{1/2}\phi \sim M_P$, resulting in a conventional type of scenario.
Therefore, intermediate scale inflation is a possible consequence of a
physical Jordan frame, due to the fact that the effective Planck mass is
also brought down to this scale. However, in a physical Einstein frame, the
toy model intermediate scale inflation is not realized.

\section{Summary and Concluding Remarks}

Physically motivated scalar-tensor theories of gravity can arise in various
contexts. We have considered a general form of scalar-tensor theory for a
single scalar field $\phi $, with no external sources. The theory is
characterized by three arbitrary functions describing the nonminimal
gravitational coupling of the scalar $F(\phi )$, the scalar kinetic function 
$k(\phi )$, and the scalar potential $V(\phi )$. Upon adopting a flat
Robertson-Walker metric, the cosmological equations of motion were obtained.
The usual slow roll conditions for inflation and the slow roll parameters
have been generalized to accommodate the arbitrariness of the functions $F$, 
$k$, and $V$. The generality of these results allows them to be applicable
to various models, including generalized Brans-Dicke theory and dilaton
gravity. The dependence of the generalized slow roll parameters upon the
functions $F$, $k$, and $V$ (or the ``effective potential'' $U=V/F^2$) shows
that several aspects of inflation, such as the onset and the end of
inflation, the amount of inflation, and the inflationary solutions
themselves, are controlled by more than just the scalar potential, and that
scalar-tensor theory can therefore introduce nontrivial modifications to
inflationary scenarios based upon minimally coupled models.

Next, the possibility of inflation occurring at an intermediate scale, for
which $\phi $ takes values well below the usual Planck scale, was
investigated by using a simple toy model where the coupling function $F$ is
a decreasing function of $\phi $, leading to an increase in the strength of
gravity at intermediate scales. The model has a standard type of quartic
Higgs potential with a low energy symmetry breaking scale $\eta $, with $%
\eta \approx 1$ TeV taken as an example. In this type of model, intermediate
scale inflation occurs for values of $\phi $ on the order of $(\eta M)^{1/2}$
(where $M=M_P/\sqrt{8\pi }=\kappa ^{-1}$), corresponding to a scale of $%
10^{10}$ -- $10^{11}$ GeV. The ``effective'' Planck scale itself is $\bar{M}%
_P=(\eta M_P)^{1/2}$, which leads to an intriguing possibility of strong
gravitational effects showing up at this scale. Although the scale of
inflation is not distanced from the effective Planck scale any more than in
the usual minimally coupled case, the model demonstrates how a low energy
symmetry breaking scale can give rise to an effective Planck scale $\bar{M}%
_P\ll M_P$. After a period of inflation in the toy model, the scalar field
evolves and eventually enters its vacuum state where $F(\eta )=1$, and the
nonminimally coupled model is approximated by a minimally coupled one (with
small corrections) for $\phi \approx \eta $. (The evolution of density
perturbations within a scalar-tensor theory is a separate problem which has
not been addressed here, and so it is not clear what constraints may be
imposed by these considerations.)

The issues considered here indicate that if reality is described by some
type of effective scalar-tensor theory with a Jordan frame realization, then
there could be interesting modifications to some of our conventional ideas
concerning relevant energy scales for high energy physics and cosmology.

\ 

\appendix 

\section{Slow Roll Equations of Motion}

Here we apply the slow roll conditions (\ref{e20})--(\ref{e22}) to reduce
the cosmological equations (\ref{e10}) and (\ref{e11}) to simpler forms. The
equations are 
\begin{equation}
H^2=\frac{\kappa ^2}{3F}\left[ \frac 12k\dot{\phi}^2+V\right] -\frac{%
F_{,\phi }}FH\dot{\phi},  \label{aa1}
\end{equation}
\begin{equation}
\begin{array}{l}
\left[ 
{\displaystyle {3(F_{,\phi })^2 \over 2F}}
+\kappa ^2k\right] (\ddot{\phi}+3H\dot{\phi})+\left\{ 
{\displaystyle {F_{,\phi } \over 2F}}
\left[ 3F_{,\phi \phi }+\kappa ^2k\right] +\frac 12\kappa ^2k_{,\phi
}\right\} \dot{\phi}^2 \\ 
+\kappa ^2F^2U_{,\phi }=0,
\end{array}
\label{aa2}
\end{equation}

\noindent  and we can drop the first term in brackets in (\ref{aa1}). The
last term on the right hand side (RHS) of this equation has a magnitude
proportional to $H|\dot{F}/F|\ll H^2$, and so it can also be dropped.
Equation (\ref{aa1}) therefore reduces to 
\begin{equation}
H^2=\frac{\kappa ^2}{3F}V.  \label{a1}
\end{equation}

For the second equation, we can first drop the acceleration term $\ddot{\phi}
$. Now let us first focus upon the first two terms on the left hand side
(LHS). Although we can not directly compare the relative magnitudes of the
terms $%
{\displaystyle {3(F_{,\phi })^2 \over 2F}}
$ and $\kappa ^2k$, we can see that the first term, proportional to $\left[ 
{\displaystyle {3(F_{,\phi })^2 \over 2F}}
\right] (3H\dot{\phi})=\frac 32F_{,\phi }(3H\dot{F}/F)\ll H^2F_{,\phi }$ is
negligible in comparison to the last term on the RHS, $\kappa ^2F^2U_{,\phi
} $. To see this, we write out this term, 
\begin{equation}
\kappa ^2F^2U_{,\phi }=\kappa ^2V_{,\phi }-\kappa ^2F_{,\phi }\frac VF%
=\kappa ^2V_{,\phi }-F_{,\phi }(3H^2),  \label{a2}
\end{equation}

\noindent which contains a part on the order of $H^2F_{,\phi }$. Equation (%
\ref{aa2}) now has reduced to 
\begin{equation}
\left\{ 
{\displaystyle {F_{,\phi } \over 2F}}
\left[ 3F_{,\phi \phi }+\kappa ^2k\right] +\frac 12\kappa ^2k_{,\phi
}\right\} \dot{\phi}^2+\kappa ^2k(3H\dot{\phi})+\kappa ^2F^2U_{,\phi }=0.
\label{a3}
\end{equation}

\noindent The entire first term on the LHS of this equation can now be
dropped. This can be seen by looking at the magnitudes of each piece: 
\[
\begin{array}{l}
\frac 32\left| F_{,\phi }(F_{,\phi \phi }/F)\right| \dot{\phi}^2=\frac 32%
\left| (F_{,\phi }/F)(\partial F_{,\phi }/\partial t)\dot{\phi}\right| =%
\frac 32\left| 
{\displaystyle {\dot{F}(\partial F_{,\phi }/\partial t) \over F}}
\right| \ll \frac 32H^2|F_{,\phi }| \\ 
\left| 
{\displaystyle {\kappa ^2F_{,\phi } \over 2F}}
k\right| \dot{\phi}^2=\frac 12\kappa ^2\left| 
{\displaystyle {\dot{F} \over F}}
k\dot{\phi}\right| \ll \frac 12\kappa ^2H\left| k\dot{\phi}\right| \\ 
\frac 12\kappa ^2\left| k_{,\phi }\right| \dot{\phi}^2=\frac 12\kappa
^2\left| \dot{k}\dot{\phi}\right| \ll \frac 12\kappa ^2H\left| k\dot{\phi}%
\right|
\end{array}
\]

\noindent Each of these pieces is dominated by remaining terms in (\ref{a3})
and are therefore discarded. Equation (\ref{aa2}) therefore reduces to the
slow roll EOM 
\begin{equation}
3Hk\dot{\phi}+F^2U_{,\phi }=0.  \label{a4}
\end{equation}

\smallskip\ 

\smallskip\ 

\

\end{document}